\documentclass[10pt]{article}
\usepackage{amsmath, latexsym, amssymb, amscd, amsfonts, amsthm, epsfig, mathrsfs, qcircuit, graphicx, url}
\usepackage[english]{babel}

\DeclareFixedFont{\sfracFont}{U}{euf}{b}{n}{7pt}
\newtheoremstyle{mydefi}
  {15pt}
  {15pt}
  {}
  {}
  {\bfseries}
  {:}
  {.5em}
  {}
\newtheoremstyle{mytheo}
  {15pt}
  {15pt}
  {\slshape}
  {}
  {\bfseries}
  {:}
  {.5em}
  {}

\theoremstyle{mytheo}
\newtheorem{theorem}{Theorem}[section]

\theoremstyle{mydefi}

\begin{document}
\title{Implementing quantum stochastic differential equations on a quantum computer}
\author{G\'e Vissers\footnote{To whom correspondence should be addressed: ge@q1t.nl}\ \ and Luc Bouten\footnote{luc@q1t.nl}\\ \\
\normalsize{Q1t BV, Lindenlaan 15, 6584 AC, Molenhoek, The Netherlands\footnote{www.q1t.nl}}}
 \date{}
\maketitle

\begin{abstract}
We study how to implement quantum
stochastic differential equations (QSDEs) on a quantum computer.
This is illustrated by an implementation of the QSDE that couples a laser
driven two-level atom to the electromagnetic field in the vacuum state
on the IBMqx4 Tenerife computer \cite{IBMQ}. We compare the resulting
master equation and quantum filtering equations to existing theory.
In this way we characterize
the performance of the computer.
\end{abstract}

\section{Introduction}
In these short notes, we study a very simple
problem with a solution that is universally well-known:
spontaneous decay of a laser-driven two-level atom
which is coupled to the electromagnetic field in the vacuum state.
The techniques  that we use to discretize the quantum
stochastic differential equation \cite{HuP84} that
describes the interaction between the two-level atom and
the laser field are very well-known
\cite{Kum85, Par88, LiP88, AtP06, BvH08, BHJ09}.
The discretized model consists of a repeated unitary
interaction of the two-level atom with subsequent field
qubits parametrized by a discretization parameter $\lambda$. The repeated interaction model easily leads
to a quantum stochastic difference equation that has
the QSDE we wish to simulate as its limit as $\lambda$ goes to zero.
Note that unitarity of the interaction is preserved in the
discretized model, which is a very desirable feature: e.g.\ after time evolution probabilities
will still always take values between $0$ and $1$. Furthermore,
the unitarity allows us to easily map the interaction on unitary gates in
a quantum computer.

The motivation for the work we present here is twofold and
aimed at a future with computers with more and more reliable
qubits:
\begin{enumerate}
\item We wish to emphasize that
quantum optics might prove a very fruitful field
of application for early quantum computers.
It is well-known how to discretize the type of
problems that arise in quantum optics
and the resulting quantum stochastic difference equations
are easy to implement on a quantum computer.
Moreover, the field of quantum optics historically
contains many interesting problems and techniques that
can serve as interesting benchmark problems for
early quantum computers.
\item On a quantum computer we can do a fully
coherent simulation of a system in interaction with
the electromagnetic field. That is, on a large enough
quantum computer, we can simulate
the complete unitary that describes the interaction between system and field, putting us
past standard analyses using master equations or
quantum filtering equations \cite{Bel92b, Car93, Dav69, BHJ07} because we also have a complete
description of the field to our availability. This could be
very useful if we wish to simulate non-Markovian networks
of systems interacting at different points with the same field,
possibly containing fully coherent feedback loops \cite{GoJ09, GoJ09b}.
\end{enumerate}

The remainder of these notes is organized as follows.
Section \ref{sec difference eq} introduces the QSDE
that we wish to simulate on a quantum computer: a laser driven
two-level atom in interaction with the vacuum EM-field.  We discuss
repeated unitary interaction models, the resulting
quantum stochastic difference equations and discuss how
to take the limit to obtain quantum stochastic differential equations
\cite{Par88, LiP88, AtP06, BvH08, BHJ09}. Next, we
introduce the repeated interaction that leads to the
QSDE corresponding to the problem at hand: a laser
driven two-level atom in interaction with the vacuum
EM-field. Section \ref{sec quantum circuit} describes
how we have implemented the repeated interaction
model of Section \ref{sec difference eq} on the
IBMqx4 Tenerife quantum computer \cite{IBMQ}.

Section \ref{sec results} presents the results of
our simulations. We compute the reduced dynamics
of the two-level atom from the simulation results and
compare it to the dynamics given by the theoretical master
equation, derived from the underlying discrete model. We
also compute the conditional dynamics of the two-level atom
conditioned on both counting photons in the field and observing a field quadrature
in the field and compare the results to the quantum filtering
equations  derived from the underlying discrete
model \cite{Brun02, GS04, BHJ09, GCMC18}. Reproducing
the correct quantum filtering equations would give an indication that
the simulation also reproduces the correlation between the field
and the two-level atom correctly.

In Section \ref{sec conclusion}, we formulate some conclusions from
our results. It should be noted that currently only very limited
simulations can be done due to the small number of reliable qubits in
the quantum computers that are currently available. In a 5 qubit
machine such as the IBMqx4, we only have 4 field qubits to our availability,
severely limiting our simulation capabilities. The work in these notes
should be seen through the lens of a hoped-for strong increase in the
computing capacity in the near future.

\section{Quantum stochastic difference equations}\label{sec difference eq}

We will now first introduce the problem that we will be studying
in this paper. We consider a two-level atom with a
coupling to the electromagnetic field. We will let the
two-level atom be driven by a laser. We will not model
the laser by an additional channel in the field that is
in an coherent state, but will directly introduce the Rabi oscillations
induced by the laser field as a Hamiltonian term in the
QSDE. We will use the following
notation throughout the paper: $\sigma_x, \sigma_y$ and $\sigma_z$
are the standard Pauli matrices. Furthermore, $\sigma_+$ and $\sigma_-$
are the \emph{raising} and \emph{lowering operator}
matrix of the two-level atom, respectively.
The interaction of the laser driven two-level atom with
the vacuum electromagnetic field is given by the
following quantum stochastic differential
equation (QSDE) in the sense of \cite{HuP84}
\begin{equation}\begin{split}\label{eq QSDE}
dU_t = \Big\{\sqrt{\kappa}\sigma_- dA_t^* - \sqrt{\kappa}\sigma_+ dA_t -\frac{1}{2}\kappa \sigma_+\sigma_- dt
- i\omega \sigma_+ \sigma_- dt - i\frac{\Omega}{2} \sigma_y dt \Big\}U_t, \ \ \ U_0 = I,
\end{split}\end{equation}
where $\kappa$ is the \emph{decay rate}, $\omega$ is the \emph{transition frequency} of the two-level
atom and $\Omega$ is the frequency of the \emph{Rabi oscillations} induced by a laser field.

In order to simulate Eqn \eqref{eq QSDE} on a quantum
computer, we first need to introduce the discrete
models (see e.g.\ \cite{BHJ09} for a detailed introduction) that
in a suitable limit will converge to a quantum stochastic differential
equation \cite{Par88, LiP88, AtP06, BvH08} such as Eqn \eqref{eq QSDE}.
To this end we first define
a time interval $[0,T]$. We divide this time interval into
$N$ equal sub intervals of length $T/N$.
We define $\lambda := \sqrt{T/N}$, i.e.\
we have $N$ sub intervals of length $\lambda^2$.
With each sub interval we associate a two-level quantum
system representing the slice of the (truncated) field that interacts
with the two-level atom at that moment.
In this way we obtain a repeated interaction
\begin{equation}\label{eq repeated interaction}
U(l)  = \overleftarrow{\prod}_{i = 1}^l M_i = M_l M_{l-1}\ldots M_2M_1, \ \ \ \ 1 \le l \le N.
\end{equation}
Here $M_i$ is a unitary operator that couples
the two-level atom and the $i$th field slice which is here also
represented by a two-level system. Note that we take all $M_i$'s to be
identical apart from the fact that they all act on their own
slice of the field. Furthermore, we will let the $M_i$'s
be a function of $\lambda$, such that if $\lambda$
goes to $0$ (i.e.\ $N$ goes to infinity) the $M_i$'s converge
to the identity map $I$. That is, we will get more and
more interactions, with smaller and smaller effect.

We now introduce linear operators
$M^\pm, M^+, M^-$ and $M^0$ \cite{BHJ09}, acting on the
two-level atom Hilbert space, in such a way that we have the following
decomposition
\begin{equation}\label{eq decomposition}
M_i - I =  M^\pm \otimes \Delta \Lambda(i) + M^+\otimes \Delta A^*(i) + M^- \otimes \Delta A(i) + M^0 \otimes \Delta t(i),
\end{equation}
where the \emph{discrete quantum noises} (see e.g. \cite{BHJ09}) are given by
\begin{equation*}\begin{split}
&\Delta \Lambda(i) :=(\sigma_+ \sigma_-)_i, \ \ \ \
 \Delta A^*(i) := \lambda(\sigma_+)_i, \\
&\Delta A(i) := \lambda(\sigma_-)_i,\ \ \ \
 \Delta t(i) := \lambda^2 I_i.
\end{split}\end{equation*}
Note that this decomposition uniquely defines
the coefficients $M^\pm, M^+, M^-$ and $M^0$
and note that these coefficients are all a function of $\lambda$,
which we leave implicit to keep our notation light. Furthermore, we usually
omit the tensor products in Eqn \eqref{eq decomposition}
to keep the notation light.

We can now write Eqn \eqref{eq repeated interaction} as the
following \emph{quantum stochastic difference equation} (see e.g. \cite{BHJ09})
\begin{equation}\begin{split}\label{eq difference equation}
\Delta U(l) :={ }&U(l)-U(l-1) = \\
={ }&\Big\{M^\pm \Delta \Lambda(l) +
M^+\Delta A^*(l) + M^- \Delta A(l) + M^0 \Delta t(l)\Big\}U(l-1),
\end{split}\end{equation}
where $1 \le l \le N$ and $U(0) = I$. We now have the following theorem
due to Parthasarathy \cite{Par88} (weak convergence), Parthasarathy and Lindsay \cite{LiP88} (weak convergence of the quantum flow) and
Attal and Pautrat \cite{AtP06} (strong convergence uniform on compact time intervals).
We state the theorem without giving the precise meaning of the mode of convergence
of the repeated interaction model to the unitary solution of the QSDE
because we would need to introduce further mathematical
details that would make us stray too far from the main narrative of
this article (see however \cite{AtP06}, or \cite{BvH08}).

\begin{theorem}\label{thm main thm} {\bf (Parthasarathy, Lindsay, Attal and Pautrat \cite{Par88, LiP88, AtP06})}
Suppose the
following limits exist:
\begin{equation}\begin{split}\label{eq definition}
&S := \lim_{\lambda \to 0} M^\pm + I, \ \ \ \  L := \lim_{\lambda \to 0} M^+ \\
&L^\dag :=   \lim_{\lambda \to 0} - M^-S^*, \ \ \ \ H :=  \lim_{\lambda \to 0} i M^0 + \frac{i}{2} L^* L,
\end{split}\end{equation}
then it follows that $S$ is unitary,  $L^\dag$ is the
adjoint of $L$, i.e. $L^* = L^\dag$, $H$ is self-adjoint and $U([t/\lambda^2])$
(where the brackets [ ] stand for rounding down to an integer)
converges to a unitary $U_t$ given by the following QSDE
\begin{equation}\label{eq general QSDE}
dU_t = \Big\{(S-I)d\Lambda_t + LdA_t^* - L^*SdA_t - \frac{1}{2}L^*Ldt - iHdt\Big\}U_t, \ \ \ \ U_0 = I.
\end{equation}
\end{theorem}

We proceed by guessing an interaction unitary $M_i$ in the
repeated interaction Eqn \eqref{eq repeated interaction} and check
via Thm \ref{thm main thm} that it leads to the correct limit coefficients
to reproduce  in the limit our system of interest
which is given  Eqn \eqref{eq QSDE}. Note that there are several $M_i$s that
will lead to the correct limit system. We will take the following $M_i$ and will show
that it indeed leads to Eqn \eqref{eq QSDE} in the limit:
\begin{equation}\begin{split}\label{eq Mi}
M_i{ } &= \exp\Big(\sqrt{\kappa}\sigma_-\otimes \lambda\sigma_+ - \sqrt{\kappa}\sigma_+\otimes \lambda\sigma_-\Big)
          \exp\Big( -i \omega \sigma_+\sigma_-\otimes  \lambda^2 I_i\Big)
          \exp\Big( -i \frac{\Omega}{2} \sigma_y \otimes  \lambda^2 I_i\Big)  \\
&= \tiny\begin{pmatrix}
             e^{ -i \omega \lambda^2}\cos(\frac{\Omega\lambda^2}{2}) &   -e^{ -i \omega \lambda^2}\sin(\frac{\Omega\lambda^2}{2})  & 0 & 0 \\
           \cos(\sqrt{\kappa}\lambda)\sin(\frac{\Omega\lambda^2}{2}) & \cos(\sqrt{\kappa}\lambda)\cos(\frac{\Omega\lambda^2}{2}) & e^{ -i \omega \lambda^2} \sin(\sqrt{\kappa}\lambda)\cos(\frac{\Omega\lambda^2}{2}) & -e^{ -i \omega \lambda^2} \sin(\sqrt{\kappa}\lambda)\sin(\frac{\Omega\lambda^2}{2}) \\
            -\sin(\sqrt{\kappa}\lambda)\sin(\frac{\Omega\lambda^2}{2})  &  -\sin(\sqrt{\kappa}\lambda)\cos(\frac{\Omega\lambda^2}{2}) &   e^{ -i \omega \lambda^2}\cos(\sqrt{\kappa}\lambda)\cos(\frac{\Omega\lambda^2}{2}) & -e^{ -i \omega \lambda^2}\cos(\sqrt{\kappa}\lambda)\sin(\frac{\Omega\lambda^2}{2}) \\
           0 & 0 & \sin(\frac{\Omega\lambda^2}{2}) & \cos(\frac{\Omega\lambda^2}{2}) \\
           \end{pmatrix}.
\end{split}\end{equation}
A short calculation then reveals
\begin{equation}\begin{split}\label{eq coefficients}
&M^0 = \begin{pmatrix}
           \frac{e^{ -i \omega \lambda^2}\cos(\sqrt{\kappa}\lambda)\cos(\frac{\Omega\lambda^2}{2})-1}{\lambda^2} & \frac{-e^{ -i \omega \lambda^2}\cos(\sqrt{\kappa}\lambda)\sin(\frac{\Omega\lambda^2}{2})}{\lambda^2}  \\
           \frac{ \sin(\frac{\Omega\lambda^2}{2})}{\lambda^2} & \frac{\cos(\frac{\Omega \lambda^2}{2})-1}{\lambda^2} \\
           \end{pmatrix}, \\
&M^- = \begin{pmatrix}
            \frac{-\sin(\sqrt{\kappa}\lambda)\sin(\frac{\Omega\lambda^2}{2})}{\lambda}   & \frac{-\sin(\sqrt{\kappa}\lambda)\cos(\frac{\Omega\lambda^2}{2})}{\lambda}\\
            0 & 0
            \end{pmatrix}, \\
& M^+ = \begin{pmatrix}
            0  &  0 \\
          \frac{e^{ -i \omega \lambda^2} \sin(\sqrt{\kappa}\lambda)\cos(\frac{\Omega\lambda^2}{2})}{\lambda} & \frac{-e^{ -i \omega \lambda^2} \sin(\sqrt{\kappa}\lambda)\sin(\frac{\Omega\lambda^2}{2})}{\lambda}
            \end{pmatrix}, \\
&M^\pm =  \begin{pmatrix}
           e^{ -i \omega \lambda^2}\big(1-\cos(\sqrt{\kappa}\lambda)\big)\cos(\frac{\Omega\lambda^2}{2}) & e^{ -i \omega \lambda^2}(\cos(\sqrt{\kappa}\lambda)-1)\sin(\frac{\Omega\lambda^2}{2})\\
           (\cos(\sqrt{\kappa}\lambda)-1)\sin(\frac{\Omega\lambda^2}{2}) & (\cos(\sqrt{\kappa}\lambda)-1)\cos(\frac{\Omega\lambda^2}{2}) \\
           \end{pmatrix}.
\end{split}\end{equation}
Using the definition of $S, L$ and $H$ in Eqn \eqref{eq definition}, we find
\begin{equation*}\begin{split}
&S =  I, \ \ \ \  L = \sqrt{\kappa} \sigma_- , \ \ \ \ H = \omega \sigma_+ \sigma_- + \frac{\Omega}{2}\sigma_y.
\end{split}\end{equation*}
That is, we have found a repeated interaction model
that converges to the QSDE of
Eqn \eqref{eq QSDE}.

\section{The quantum circuit}\label{sec quantum circuit}

The various contributions to the interaction between the atom and the field,
given in Eqn \eqref{eq Mi}, can easily be mapped to elementary quantum
gates.
The following quantum circuit \cite{NiC00} implements this interaction for a single time
slice:

\begin{center}
\makebox{\Qcircuit @C=1em @R=.7em {
     \lstick{atom}  & \gate{R_y(\Omega\lambda^2)} &
\gate{R_z(\omega\lambda^2)}
         & \ctrl{1} & \gate{R_y(2\sqrt{\kappa}\lambda)} & \ctrl{1} & \qw \\
     \lstick{field} & \qw                         & \qw
         & \targ\   & \ctrl{-1}                         & \targ    & \qw
}}
\end{center}

For modeling multiple time slices, different qubits are used to
describe the field at different times, and coupled to
the atom qubit. Currently, this limits the simulation to
a maximum of four time slices on the IBMqx4 computer.
If operations based on classical bits are enabled, we
could reuse the same qubit for the field by measuring this qubit, and rotating it back to its $| 0\rangle$
state when the outcome is $1$. Unfortunately, this is currently only
implemented in simulators and not in real hardware.

At the end of the simulation, all qubits (both field and atom) were measured.
The atom was measured in the $x$, $y$, and $z$ basis, whereas the field qubits
were measured in the $x$ and $z$ basis. Statistics were collected over 10,240 runs
for each combination of measurement directions of atom and field.

\section{Results}\label{sec results}

\subsection{Master equation}
\begin{figure}
\centering
\includegraphics[width=0.6\textwidth]{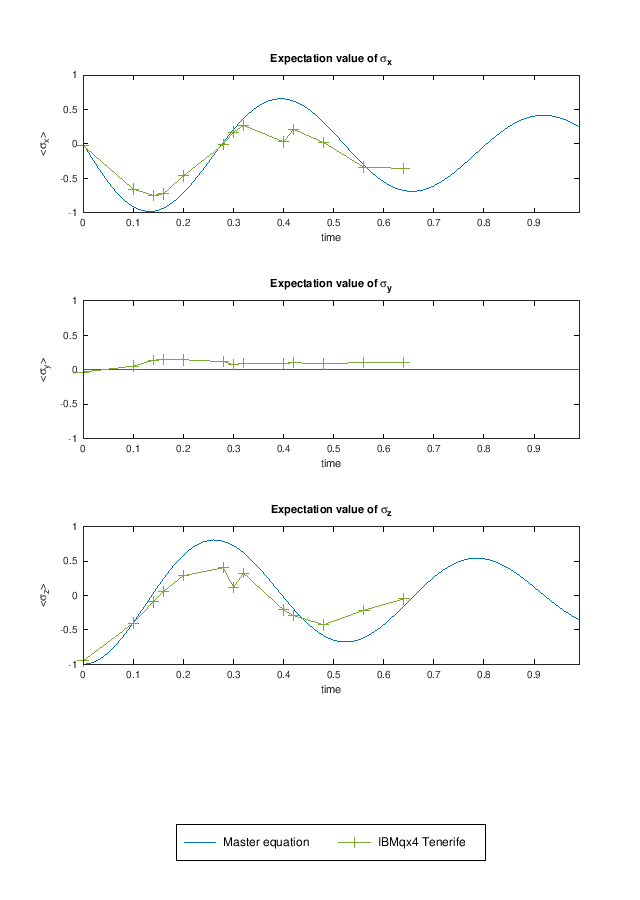}
\caption{Calculated expectation values for the Pauli spin operators for the two-level atom,
using parameters $\kappa=1$, $\omega=0$ and $\Omega=12$.
Results are plotted for the master equation Eqn \eqref{eq discrete master},
and calculations on the actual
IBMqx4 Tenerife quantum computer. The results of the master equation
are obtained using a time step $\lambda^2=0.01$. The IBMqx4 curve consists of
collected results of experiments with time steps $\lambda^2=0.16$, $0.14$, and $0.10$,
each averaged over 10240 runs.}
\label{fig master equation}
\end{figure}

The repeated interaction model given by Eqn \eqref{eq Mi}
leads to the following discrete time master equation \cite[Eqn 4.3, page 265]{BHJ09}
for the state $\rho_l$ of the two-level atom:
\begin{equation}\label{eq discrete master}
\Delta \rho_l := \rho_l - \rho_{l-1} =\mathcal{L}(\rho_{l-1})\lambda^2,\ \ \ \
\rho_0 = \begin{pmatrix} 0 & 0 \\ 0 & 1\end{pmatrix},
\end{equation}
where the discretized Lindblad operator is
given by \cite[Eqn 4.4, page 265]{BHJ09}
\begin{equation}\label{eq discrete Lindblad}
\mathcal{L}(\rho) := M^+ \rho {M^+}^* + \lambda^2 M^0 \rho {M^0}^* + M^0\rho + \rho {M^0}^*.
\end{equation}
Here $M^+$ and $M^0$ are given by Eqn \eqref{eq coefficients}.
Figure \ref{fig master equation} compares the
theoretical results given by the master
equation Eqn \eqref{eq discrete master} and the
results obtained with the IBMqx4 Tenerife quantum computer.

\subsection{Homodyne quantum filter}
\begin{figure}
\centering
\includegraphics[width=\textwidth]{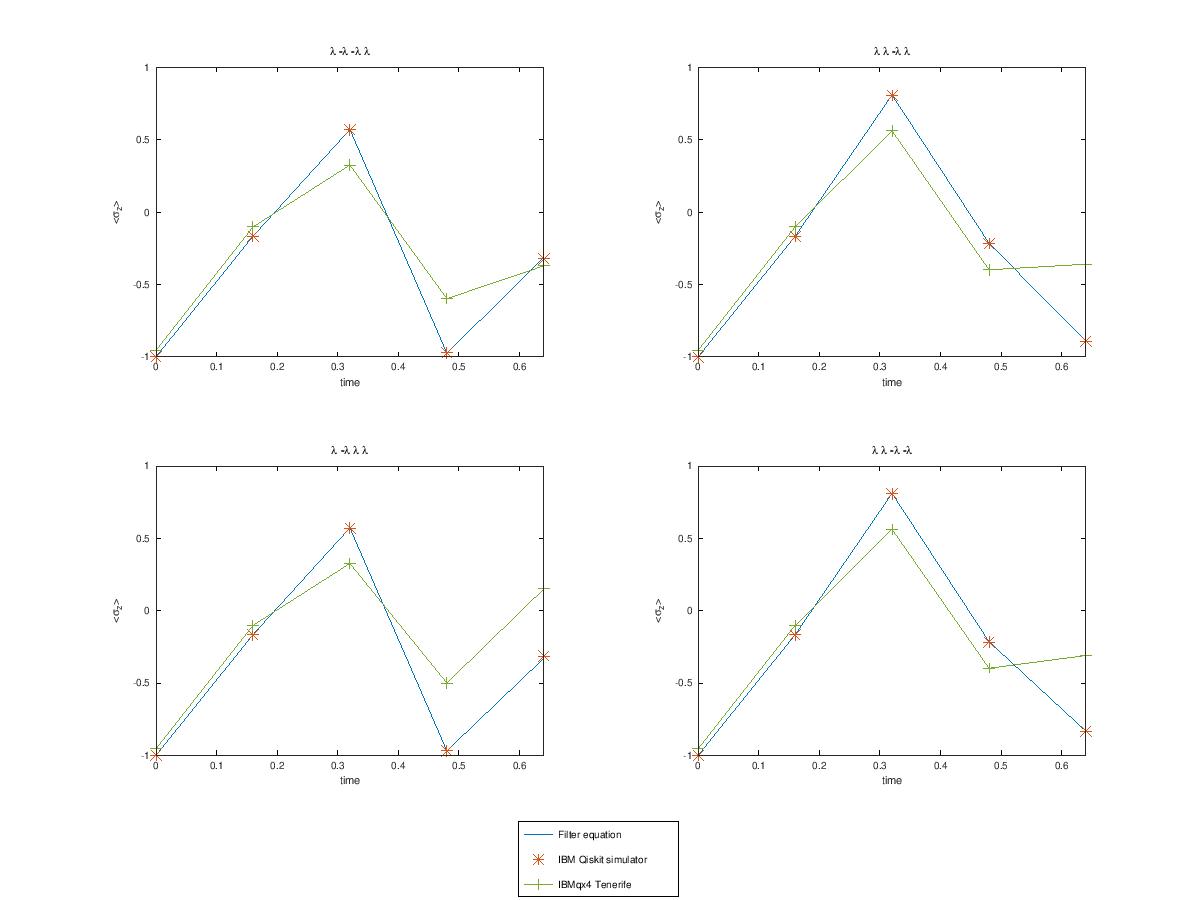}
\caption{Time evolution of the expectation value of the Pauli $\sigma_z$ operator
of the two-level atom,
in the homodyne detection scheme, for the four trajectories that have accumulated
the most statistics. The trajectory is given by the title above the plots, which shows
$\Delta Y$. The filter equation results are from
Eqn \eqref{eq discrete quantum filter homodyne}.
The IBMqx4 results are averaged over 10,240 runs. The simulator results are obtained
from the IBM Qiskit simulator, using the same quantum assembly code that was used
on the IBMqx4 Tenerife computer, and are averaged over 102,400 runs.}
\label{fig homodyne z}
\end{figure}

\begin{figure}
\centering
\includegraphics[width=\textwidth]{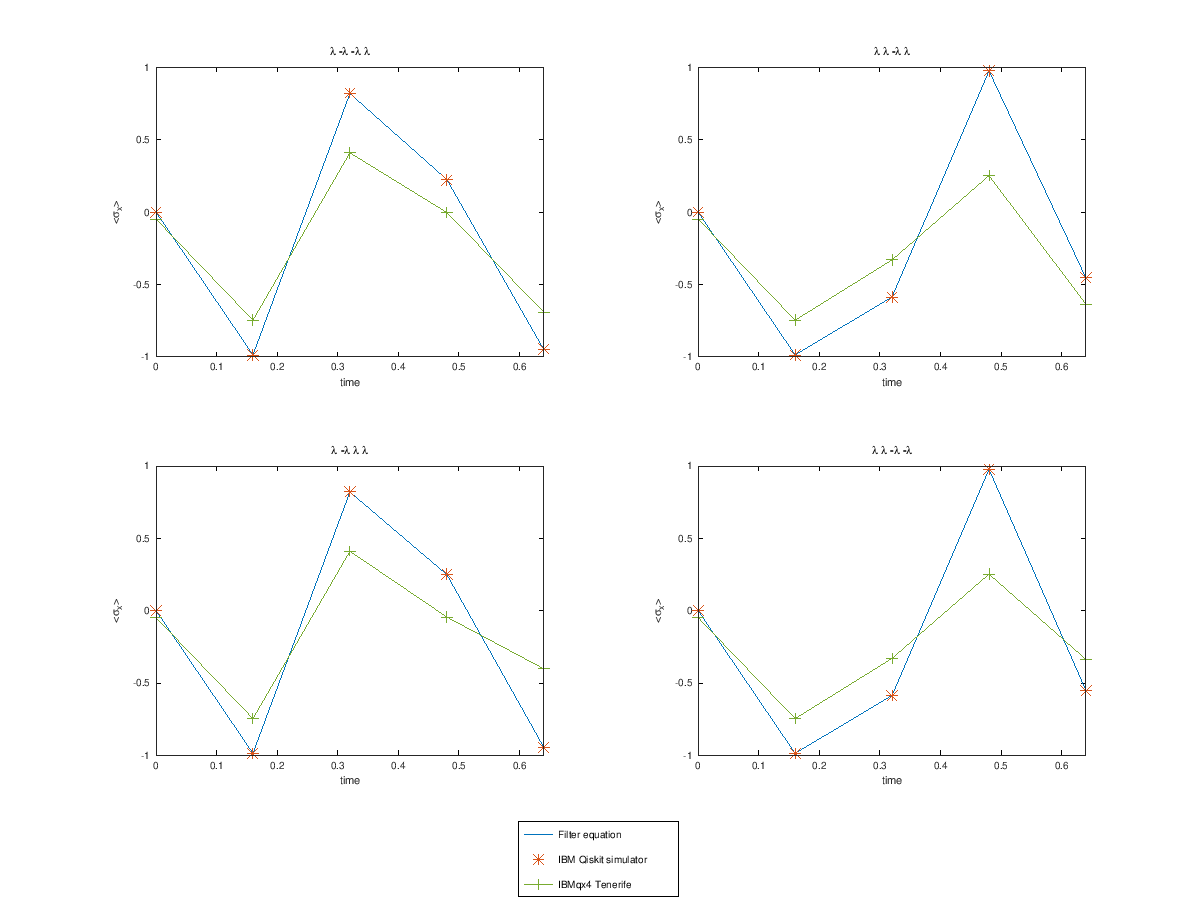}
\caption{As Figure \ref{fig homodyne z}, for the $\sigma_x$ operator.}
\label{fig homodyne x}
\end{figure}

\begin{figure}
\centering
\includegraphics[width=\textwidth]{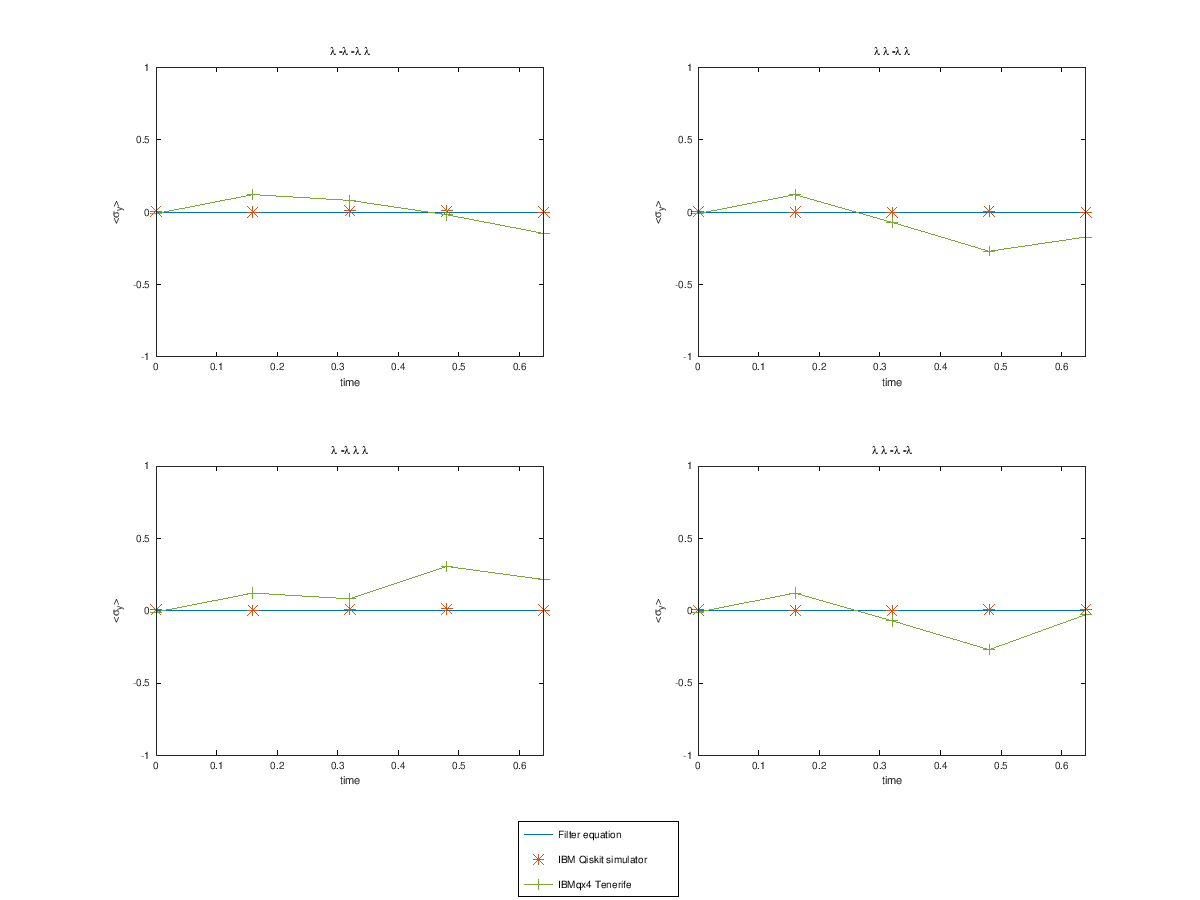}
\caption{As Figure \ref{fig homodyne z}, for the $\sigma_y$ operator.}
\label{fig homodyne y}
\end{figure}

We now turn to the situation where we are not simply
tracing over the field, but condition on observations
made in the field. Suppose that for $l = 1,2,3,4$,
we have observed $\lambda*\sigma_x$  in the field:
\begin{equation*}
\Delta Y(l) := \lambda * \mbox{"the outcome of the\ } \sigma_x \mbox{-measurement of the\ }l\mbox{th field qubit"}.
\end{equation*}
Physically, this corresponds to the case where we observe
the field with a homodyne detection setup after the interaction
with the two-level atom. We can condition the time evolution of the density
matrix on an observed homodyne photo current detection record.
The conditioned density matrix obeys the following \emph{discrete
quantum filtering equation for homodyne detection} \cite[Section 5.2, page 274]{BHJ09}
\begin{equation}\begin{split}\label{eq discrete quantum filter homodyne}
\Delta  \rho_l = \mathcal{L}(\rho_{l-1})\lambda^2 + \frac{\mathcal{J}(\rho_{l-1})- \mbox{tr}[\mathcal{J}(\rho_{l-1})] \Big( \rho_{l-1}+\lambda^2\mathcal{L}(\rho_{l-1})\Big)}
                                                                                            {1-\lambda^2 \mbox{tr}[\mathcal{J}(\rho_{l-1})]^2}
                       \Big(\Delta Y(l) - \mbox{tr}[\mathcal{J}(\rho_{l-1})]\lambda^2\Big),
\end{split}\end{equation}
where $\mathcal{L}$ is given by Eqn \eqref{eq discrete Lindblad} and $\mathcal{J}$ and the initial state $\rho_0$
are given by
\begin{equation*}
\mathcal{J}(\rho) := M^+\rho + \rho{M^+}^* + \lambda^2M^+\rho{M^0}^* + \lambda^2M^0 \rho{M^+}^*, \ \ \ \
\rho_0=  \begin{pmatrix} 0 & 0 \\ 0 & 1\end{pmatrix},
\end{equation*}
Here $M^+$ and $M^0$ are given by Eqn \eqref{eq coefficients}.
Figures \ref{fig homodyne z}, \ref{fig homodyne x} and \ref{fig homodyne y} compare the
theoretical results given by the quantum filter
equation Eqn \eqref{eq discrete quantum filter homodyne} and the
results obtained with the IBM Qiskit \cite{Qiskit} simulator and
the IBMqx4 Tenerife quantum computer.

\subsection{Counting quantum filter}
\begin{figure}
\centering
\includegraphics[width=\textwidth]{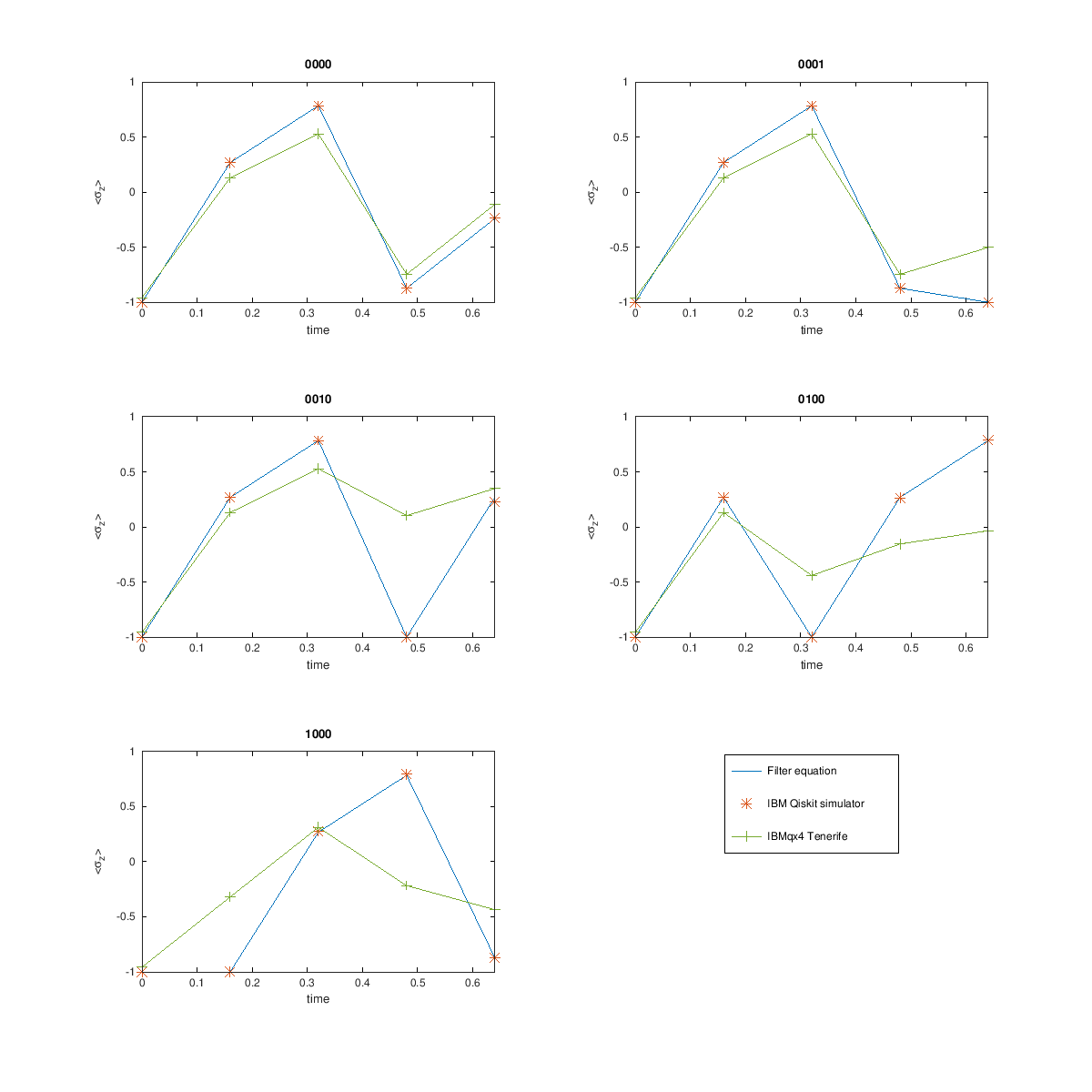}
\caption{Time evolution of the expectation value of the Pauli $\sigma_z$ operator
of the two-level atom,
in the photon counting scheme, for the case where no photons were detected (0000),
and where a single photon was detected after the fourth (0001), third (0010), second (0100)
and first (1000) time step. The filter equation results are from Eqn \eqref{eq discrete quantum filter photon counting}.
The IBMqx4 results are averaged over 10,240 runs. The simulator results are obtained
from the IBM Qiskit simulator, using the same quantum assembly code that was used
on the IBMqx4 Tenerife computer, and are averaged over 102,400 runs.}
\label{fig counting z}
\end{figure}

\begin{figure}
\centering
\includegraphics[width=\textwidth]{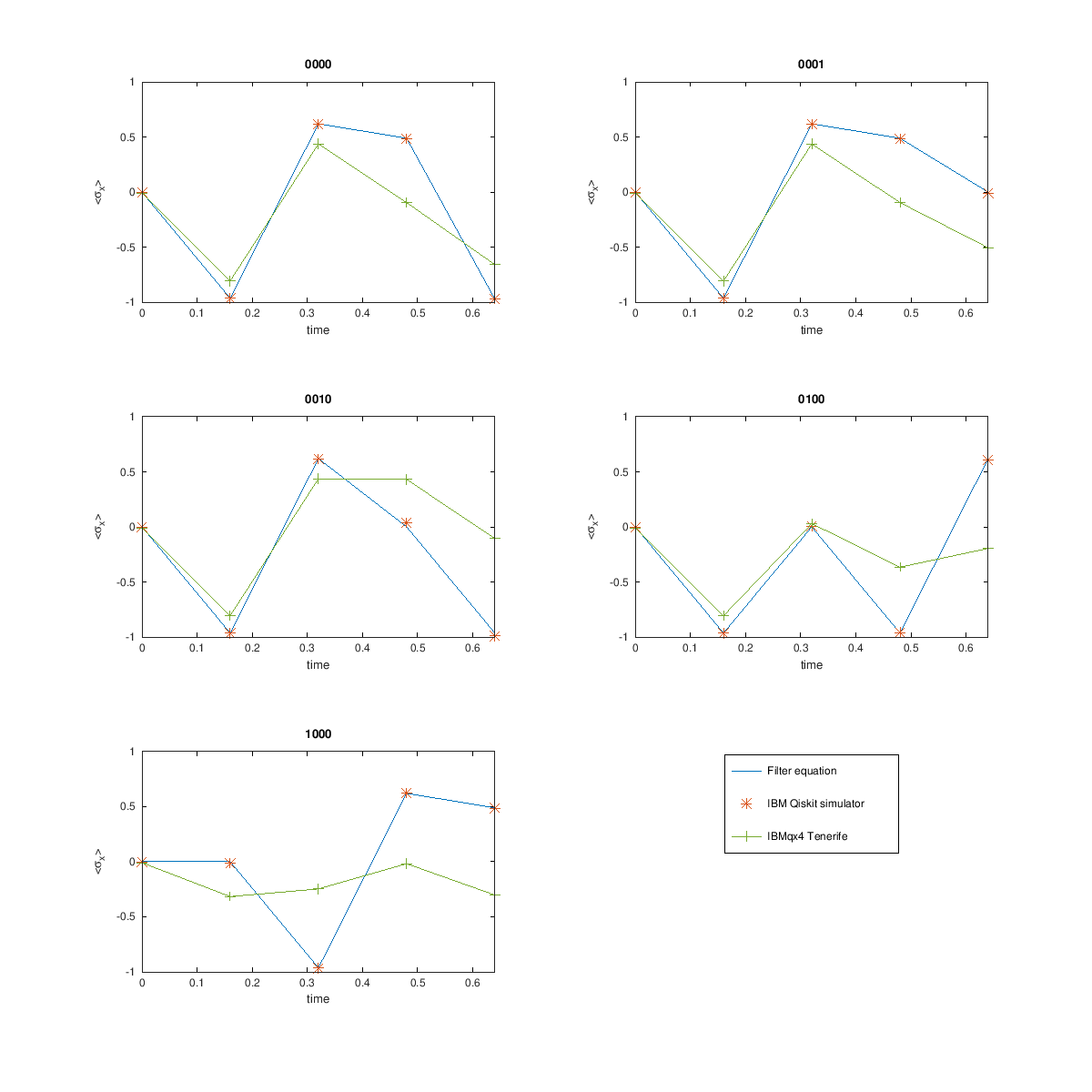}
\caption{As Figure \ref{fig counting z}, for the $\sigma_x$ operator.}
\label{fig counting x}
\end{figure}

\begin{figure}
\centering
\includegraphics[width=\textwidth]{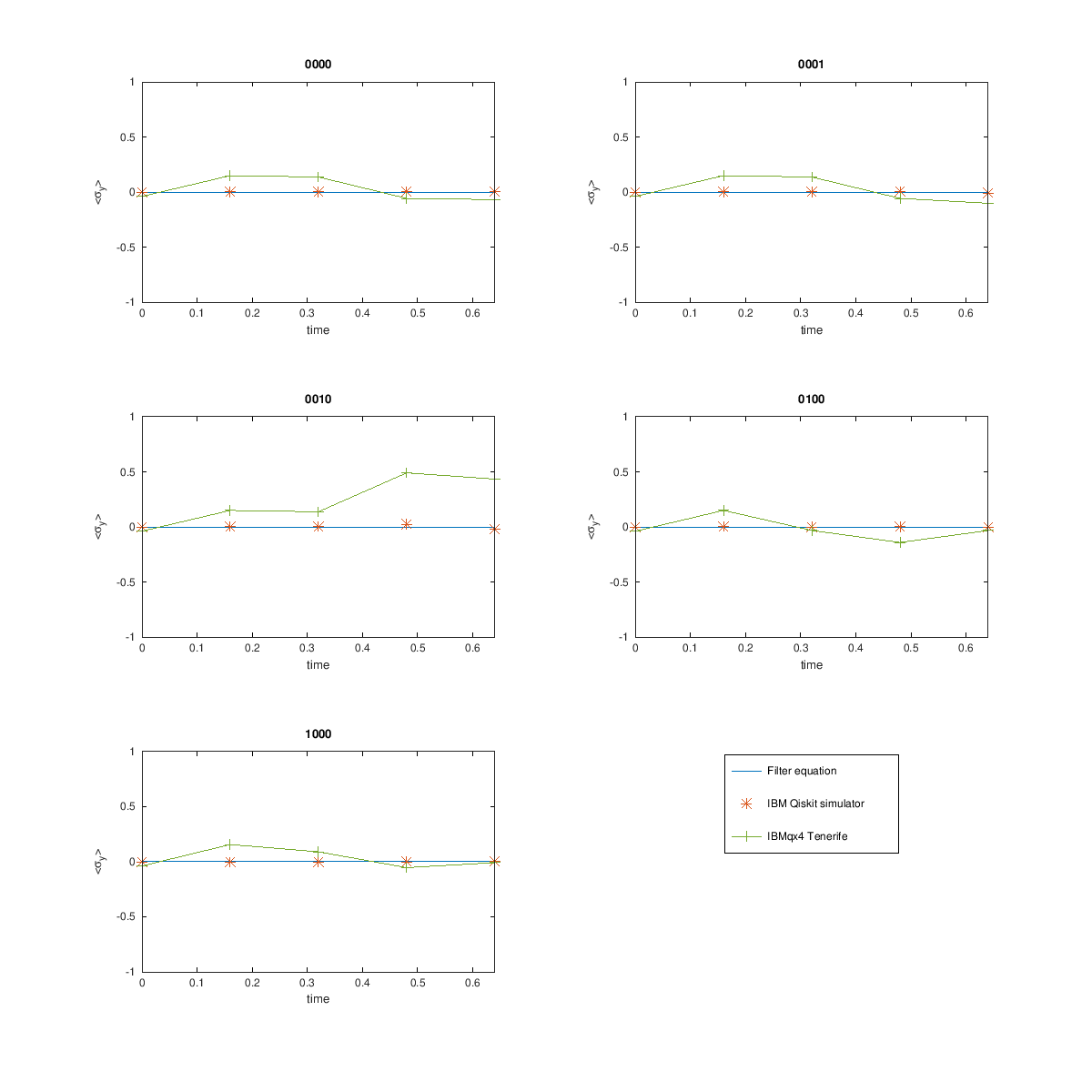}
\caption{As Figure \ref{fig counting z}, for the $\sigma_y$ operator.}
\label{fig counting y}
\end{figure}

Finally, we turn to the situation where we condition on having
observed $\sigma_+\sigma_-$ in the field for $l=1,2,3,4$.
That is, we have the following observations:
\begin{equation*}
\Delta Y(l) = \frac{1 + \mbox{"the outcome of the\ } \sigma_z \mbox{-measurement
of the\ }l\mbox{th field qubit"}}{2}.
\end{equation*}
 Physically, this corresponds to the case where we
observe the field with a photo detector after the interaction with the two-level
atom. We can condition the time evolution of the density matrix on an observed
photo detection record. The conditioned density matrix obeys the following
\emph{discrete quantum filtering equation for photon counting} \cite[Section 5.3, page 276]{BHJ09}
\begin{equation}\begin{split}\label{eq discrete quantum filter photon counting}
\Delta  \rho_l{ }&= \mathcal{L}(\rho_{l-1})\lambda^2 + \frac{\frac{M^+\rho_{l-1}{M^+}^*}{\mbox{tr}[\rho_{l-1} {M^+}^* M^+]} - \rho_{l-1}-\lambda^2\mathcal{L}(\rho_{l-1})}
                                                                                            {1-\lambda^2 \mbox{tr}[\rho_{l-1} {M^+}^* M^+]}
                       \Big(\Delta Y(l) - \mbox{tr}[\rho_{l-1} {M^+}^* M^+]\lambda^2\Big), \\
\rho_0{ }&=  \begin{pmatrix} 0 & 0 \\ 0 & 1\end{pmatrix}.
\end{split}\end{equation}
Here $M^+$ is given by Eqn \eqref{eq coefficients}.
Figures \ref{fig counting z}, \ref{fig counting x} and \ref{fig counting y} compare the
theoretical results given by the quantum filter
equation Eqn \eqref{eq discrete quantum filter photon counting} and the
results obtained with the IBM Qiskit simulator and
the IBMqx4 Tenerife quantum computer.

\section{Conclusion}\label{sec conclusion}
In these notes we have shown that it is fairly
straightforward to implement quantum stochastic
differential equations on a quantum computer.
The mathematical theory \cite{Kum85, Par88, LiP88, AtP06, BvH08, BHJ09}
behind the necessary discretization of the equations
is well worked out and easily translated into a
quantum circuit. It is possible with the very limited
capacity of the currently available quantum computers
to simulate some simple quantum optical
features described by a QSDE (e.g. a Rabi oscillation).

We have also seen that the filter equations are
to a large extent correctly reproduced on the
IBMqx4 Tenerife quantum computer. This provides
confidence that the (quantum) correlations between
the atom and the field are accounted for correctly.
This opens the door to fully coherent simulations of
systems that interact with the field at different points,
even including fully coherent feedback loops \cite{GoJ09, GoJ09b}.
Naturally, this is only possible when quantum computers
are available with more and more reliable qubits.

As can be seen in Figures \ref{fig counting z} and \ref{fig counting x},
the time evolution between counts in the photon counting scheme seems to be accounted for correctly,
however, the jump operation does not seem to be accurate to the theory.
This can be understood though: there are  relatively few jumps and there
is already quite a bit of noise on the results when the field is in the vacuum
state. This means there is an identity component in the jump operator that
leads to a deviation from the theory (in which this noise is not present, but could
be modeled).

With respect to the discrete quantum filtering equations \cite{BHJ09}, we
note that
they completely coincide with the IBM Qiskit simulator results. However,
they are much less computationally intensive and could still be used if the
number of qubits is larger than the simulator can deal with. Note, however,
that in the simulator it is also possible to recycle the field qubits after they
have been measured. This is currently not yet possible on
the real IBMQ hardware.

It will be interesting to see future quantum computers simulate more
complex benchmark problems originating from quantum optics.

\bibliography{qsde_qcomp}

\begin{thebibliography}{10}

\bibitem{Qiskit}
G.~Aleksandrowicz, T.~Alexander, P.~Barkoutsos, L.~Bello, Y.~Ben-Haim,
  D.~Bucher, F.~J. Cabrera-Hern{\'a}dez, J.~Carballo-Franquis, A.~Chen, C.-F.
  Chen, J.~M. Chow, A.~D. C{\'o}rcoles-Gonzales, A.~J. Cross, A.~Cross,
  J.~Cruz-Benito, C.~Culver, S.~D. L.~P. Gonz{\'a}lez, E.~D.~L. Torre, D.~Ding,
  E.~Dumitrescu, I.~Duran, P.~Eendebak, M.~Everitt, I.~F. Sertage, A.~Frisch,
  A.~Fuhrer, J.~Gambetta, B.~G. Gago, J.~Gomez-Mosquera, D.~Greenberg,
  I.~Hamamura, V.~Havlicek, J.~Hellmers, {\L}.~Herok, H.~Horii, S.~Hu,
  T.~Imamichi, T.~Itoko, A.~Javadi-Abhari, N.~Kanazawa, A.~Karazeev,
  K.~Krsulich, P.~Liu, Y.~Luh, Y.~Maeng, M.~Marques, F.~J.
  Mart{\'\i}n-Fern{\'a}ndez, D.~T. McClure, D.~McKay, S.~Meesala, A.~Mezzacapo,
  N.~Moll, D.~M. Rodr{\'\i}guez, G.~Nannicini, P.~Nation, P.~Ollitrault, L.~J.
  O'Riordan, H.~Paik, J.~P{\'e}rez, A.~Phan, M.~Pistoia, V.~Prutyanov,
  M.~Reuter, J.~Rice, A.~R. Davila, R.~H.~P. Rudy, M.~Ryu, N.~Sathaye,
  C.~Schnabel, E.~Schoute, K.~Setia, Y.~Shi, A.~Silva, Y.~Siraichi,
  S.~Sivarajah, J.~A. Smolin, M.~Soeken, H.~Takahashi, I.~Tavernelli,
  C.~Taylor, P.~Taylour, K.~Trabing, M.~Treinish, W.~Turner, D.~Vogt-Lee,
  C.~Vuillot, J.~A. Wildstrom, J.~Wilson, E.~Winston, C.~Wood, S.~Wood,
  S.~W{\"o}rner, I.~Y. Akhalwaya, and C.~Zoufal.
\newblock Qiskit: An open-source framework for quantum computing, 2019.

\bibitem{AtP06}
S.~Attal and Y.~Pautrat.
\newblock From repeated to continuous quantum interactions.
\newblock {\em Annales~Henri~Poincar\'e}, 7:59--104, 2006.

\bibitem{Bel92b}
V.~P. Belavkin.
\newblock Quantum stochastic calculus and quantum nonlinear filtering.
\newblock {\em J. Multivar. Anal.}, 42:171--201, 1992.

\bibitem{BHJ07}
L.~Bouten, R.~{van Handel}, and M.~James.
\newblock An introduction to quantum filtering.
\newblock {\em SIAM J.~Control Optim.}, 46:2199--2241, 2007.

\bibitem{BvH08}
L.~M. Bouten and R.~{van Handel}.
\newblock Discrete approximation of quantum stochastic models.
\newblock {\em J. Math. Phys.}, 49:102109, 2008.

\bibitem{BHJ09}
L.~M. Bouten, R.~{van Handel}, and M.~James.
\newblock A discrete invitation to quantum filtering and feedback control.
\newblock {\em SIAM Review}, 51:239--316, 2009.

\bibitem{Brun02}
T.~A. Brun.
\newblock A simple model of quantum trajectories.
\newblock {\em Am. J. Phys.}, 70:719--737, 2002.

\bibitem{Car93}
H.~J. Carmichael.
\newblock {\em An Open Systems Approach to Quantum Optics}.
\newblock Springer-Verlag, Berlin Heidelberg New-York, 1993.

\bibitem{Dav69}
E.~B. Davies.
\newblock Quantum stochastic processes.
\newblock {\em Commun.~Math.~Phys.}, 15:277--304, 1969.

\bibitem{GoJ09b}
J.~Gough and M.~James.
\newblock Quantum feedback networks: Hamiltonian formulation.
\newblock {\em Commun. Math. Phys.}, 287:1109--1132, 2009.

\bibitem{GoJ09}
J.~Gough and M.~James.
\newblock The series product and its application to quantum feedforward and
  feedback networks.
\newblock {\em IEEE Trans. Aut. Control}, 54:2530--2544, 2009.

\bibitem{GS04}
J.~Gough and A.~Sobolev.
\newblock Stochastic {S}chr\"odinger equations as limit of discrete filtering.
\newblock {\em Open Syst. Inf. Dyn.}, 11:235--255, 2004.

\bibitem{GCMC18}
J.~A. Gross, C.~M. Caves, G.~J. Milburn, and J.~Combes.
\newblock Qubit models of weak continuous measurements: markovian conditional
  and open-system dynamics.
\newblock {\em Quantum Science and Technology}, 3(2):024005, 2018.

\bibitem{HuP84}
R.~L. Hudson and K.~R. Parthasarathy.
\newblock Quantum {It\^o's} formula and stochastic evolutions.
\newblock {\em Commun. Math. Phys.}, 93:301--323, 1984.

\bibitem{IBMQ}
IBM.
\newblock The {IBM Q} experience.
\newblock \url{https://quantumexperience.ng.bluemix.net/qx}.
\newblock Accessed 23 November 2018.

\bibitem{Kum85}
B.~K\"ummerer.
\newblock Markov dilations on {$W^*$-algebras}.
\newblock {\em J. Funct. Anal.}, 63:139--177, 1985.

\bibitem{LiP88}
J.~M. Lindsay and K.~R. Parthasarathy.
\newblock The passage from random walk to diffusion in quantum probability ii.
\newblock {\em Sankhya: The Indian journal of statistics}, 50:151--170, 1988.

\bibitem{NiC00}
M.~Nielsen and I.~Chuang.
\newblock {\em Quantum Computation and Quantum Information}.
\newblock Cambridge University Press, 2000.

\bibitem{Par88}
K.~R. Parthasarathy.
\newblock The passage from random walk to diffusion in quantum probability.
\newblock {\em Journal of applied probability}, special volume 25A:151--166,
  1988.

\end{thebibliography}
\end{document}